\documentclass[prb,aps,twocolumn,superscriptaddress,showpacs]{revtex4-1}

\usepackage{graphicx}
\usepackage{epstopdf}

\begin{document}

\title{Spin fluctuations and superconductivity in powders of Fe$_{1+x}$Te$_{0.7}$Se$_{0.3}$ as a function of interstitial iron concentration}

\author{C. Stock}
\affiliation{NIST Center for Neutron Research, 100 Bureau Drive, Gaithersburg, Maryland 20899, USA}
\affiliation{Indiana University, 2401 Milo B. Sampson Lane, Bloomington, Indiana 47404, USA}
\author{E. E. Rodriguez} 
\affiliation{NIST Center for Neutron Research, 100 Bureau Drive, Gaithersburg, Maryland 20899, USA}
\author{M. A. Green}
\affiliation{NIST Center for Neutron Research, 100 Bureau Drive, Gaithersburg, Maryland 20899, USA}
\affiliation{Department of Materials Science and Engineering, University of Maryland, College Park,
MD 20742}

\date{\today}

\begin{abstract}

Using neutron inelastic scattering, we investigate the role of interstitial iron on the low-energy spin fluctuations in powder samples of Fe$_{1+x}$Te$_{0.7}$Se$_{0.3}$.   We demonstrate how combining the principle of detailed balance along with measurements at several temperatures allows us to subtract both temperature-independent and phonon backgrounds from $S(Q,\omega)$ to obtain purely magnetic scattering.  For small values of interstitial iron ($x=0.009(3)$), the sample is superconducting (T$_{c}$=14 K)  and displays a spin gap of  7 meV peaked in momentum at wave vector ${\bf{q_{0}}}$=($\pi$,$\pi$) consistent with single crystal results.  On populating the interstitial iron sites, the superconducting volume fraction decreases and we observe a filling in of the low-energy magnetic fluctuations and a decrease of the characteristic wave vector of the magnetic fluctuations. For large concentrations of interstitial iron ($x$=0.048(2)) where the superconducting volume fraction is minimal, we observe the presence of gapless spin fluctuations at a wavevector of ${\bf{q_{0}}}$=($\pi$,0).    We estimate the absolute total moment for the various samples and find that the amount of interstitial iron does not change the total magnetic spectral weight significantly, but rather has the effect of shifting the spectral weight in $Q$ and energy.  These results show that the superconducting and magnetic properties can be tuned by doping small amounts of iron and are suggestive that interstitial iron concentration is also a controlling dopant in the Fe$_{1+x}$Te$_{1-y}$Se$_{y}$ phase diagram in addition to the Te/Se ratio. 

\end{abstract}


\maketitle

\section{Introduction}

The discovery of high temperature superconductivity in iron based systems has resulted in a wide variety of studies on seemingly different compounds.~\cite{Johnston10:59,Paglione10:6}  All iron based superconductors are based on two-dimensional planes of magnetic iron and initial investigations on powder samples of Ba$_{0.6}$K$_{0.4}$Fe$_{2}$As$_{2}$ found strong coupling between the magnetism and superconductivity as evidenced by the observation of a magnetic resonance peak in the superconducting phase.~\cite{Christianson08:456}    While most studies have focused on FeAs systems, presumably because of the relatively high superconducting transition temperature, Fe$_{1+x}$Te$_{1-y}$Se$_{y}$ is arguably the simplest iron based superconductor and is built on monolayers of magnetic FeTe planes bound by Van der Waals forces.~\cite{Wen11:xx}   An optimal superconducting transition temperature of 14 K has been reported for  Fe$_{1+x}$Te$_{0.5}$Se$_{0.5}$,~\cite{Yeh08:78, Sales09:79} and nearly stiochiometric FeSe having a transition temperature of 8 K.~\cite{Kotegawa08:77, McQueen09:79}

The magnetic structure of the parent non superconducting Fe$_{1+x}$Te has been investigated in powders and single crystals using neutron diffraction and has reported the existence of a commensurate double stripe spin density wave phase for small concentrations of $x$ with an ordering wave vector of ${\bf{q}_{0}}=(\frac{1}{2},0,\frac{1}{2})$.~\cite{S-Li09:79,Bao09:102,Robler11:84}  For larger concentrations of interstitial iron, the magnetic phase becomes incommensurate along the $a^{*}$ direction and the structure is believed to be defined by a magnetic spiral.  At a concentration of $x\approx0.12$ an unusual incommensurate phase was found with long-range magnetic correlations along the $c$-axis and short-range correlations within the $a-b$ plane.~\cite{Rodriguez11:84}  For superconducting concentrations of Fe$_{1+x}$Te$_{1-y}$Se$_{y}$, the magnetic order is found to be replaced by short range magnetic correlations peaked near ${\bf{q}_{0}}=(\frac{1}{2},\frac{1}{2},L)$.  Therefore, the magnetic correlations shift from the $(\pi,0)$ position to the $(\pi,\pi)$ points on becoming superconducting and with selenium doping.~\cite{Liu10:9}    The $(\pi,\pi)$ point is believed to correspond to a Fermi surface nesting wave vector measured using ARPES.~\cite{Xia09:103}

The spin fluctuations have been investigated for Fe$_{1+x}$Te$_{1-y}$Se$_{y}$ with superconducting concentrations of y$\approx$ 0.3 and 0.5.  The low energy dynamics are dominated by a resonance peak which is located near ${\bf{q_{0}}}$=($\frac{1}{2}$,$\frac{1}{2}$) and forms a rod of scattering along L, indicative of strong two dimensional fluctuations.~\cite{Qiu09:103,Argyriou10:81}  The resonance peak is onset at the superconducting transition and is located near $\approx$ 7 meV.  This energy scale is widely believed to be directly related to superconductivity as it scales with the superconducting transition temperature in iron based samples where a resonance peak has been observed.  

\begin{figure}
\includegraphics[width=8.5cm]{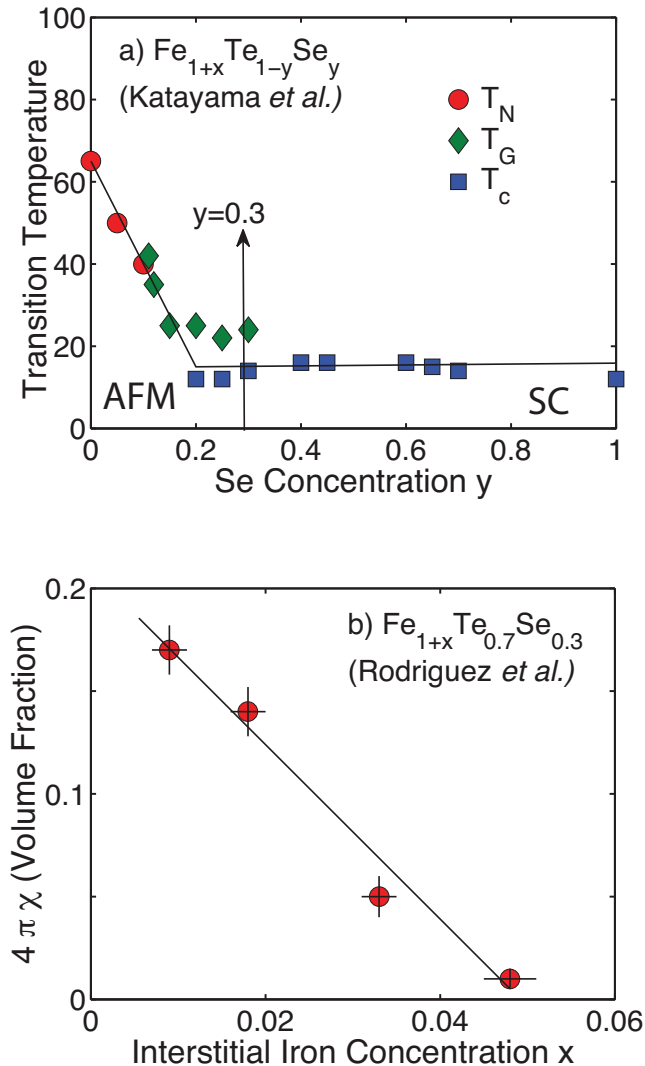}
\caption{\label{phase_diagram} (Color online) $a)$ illustrates the phase diagram derived and presented in Ref. \onlinecite{Katayama10:79}.  $b)$ plots the superconducting volume fraction as a function of interstitial iron concentration for a fixed $y$=0.3.}
\end{figure}

The  Fe$_{1+x}$Te$_{1-y}$Se$_{y}$ phase diagram representing a competition between superconductivity and antiferromagnetism is illustrated in Fig \ref{phase_diagram} $a)$ taken from Ref. \onlinecite{Katayama10:79}.   The phase diagram demonstrates that for small values of Se, antiferromagnetic order dominates while at higher concentrations, antiferromagnetism is replaced by a superconducting ground state with a relatively large transition temperature.   While the phase diagram might have some similarities to other magnetic superconductors such as the cuprates and analogous FeAs systems, there are several noteworthy differences.  First, the system is metallic for all concentrations of Se and second, the superconducting transition temperature remains fairly constant and is not tuned over the extreme range observed in other systems like FeAs and cuprate based superconductors.  There also exists a large Se concentration range where both superconductivity and antiferromagnetism is present in the same sample.~\cite{Khasanov09:80} These difference might indicate that Se is not the controlling dopant in this system and that another parameter might be present which is tuning the superconductivity.

An important avenue to explore is the effects of interstitial iron on the magnetic and electronic phase diagram of Fe$_{1+x}$Te$_{1-y}$Se$_{y}$.  Studies on Fe$_{1+x}$Te$_{1-y}$S$_{y}$ powders and single crystals have found that the cation concentration was directly tied with the anion substitution by sulphur.~\cite{Zajdel10:132, Hu09:80}   A similar relationship can be seen in single crystal studies of Fe$_{1+x}$Te$_{1-y}$Se$_{y}$.\cite{Sales09:79}  Several theoretical studies have suggested that interstitial iron has a dramatic effect on the electronic properties either altering the crystal field environment of the in-plane Fe sites, or changing the band structure.~\cite{Turner09:80,Zhang09:79}  The excess iron resides in regions between the weakly bonded layers of FeTe$_4$ tetrahedra and several studies have found that for a fixed concentration of Se, both the superconductivity and magnetism could be tuned with the concentration of this interstitial iron.~\cite{Rodriguez11:2}   Neutron inelastic scattering studies of the parent Fe$_{1+x}$Te compound have found significant effects on the low-energy magnetic fluctuations, magnetic and crystalline structure with doping interstitial iron.~\cite{Rodriguez11:84}  In particular, for small amounts of interstitial iron where magnetic order is observed at ${\bf{q_{0}}}$=($\frac{1}{2}$, 0, $\frac{1}{2}$), a large spin gap is observed at $\approx$ 7 meV whereas for large interstitial iron concentrations where incommensurate order is observed, the excitations are gapless.~\cite{Stock11:84}    Studies at the boundary between the collinear and spiral phases reported both a $\sim$ 7 meV gapped excitation and low-energy gapless incommensurate fluctuations.~\cite{Zaliznyak11:107}  Therefore, similar to the charge doping found in iron arsenide superconductors, the magnetic energy scale in the Fe$_{1+x}$Te system can also be similarly tuned with charge doping via interstitial iron.

We investigate the effects of interstitial iron on the magnetic excitations in powder samples of Fe$_{1+x}$Te$_{0.7}$Se$_{0.3}$.  The location of this study is illustrated by the vertical arrow in Fig. \ref{phase_diagram} (a) and has previously been reported to be both a superconductor and display magnetic spin-glass dynamics.\cite{Katayama10:79}   By tuning $x$ from 0 to $\approx$ 0.05 the superconducting volume fraction can be tuned to nearly 0 (Fig. \ref{phase_diagram} (b)).  We will show that magnetic fluctuations evolve from gapped excitations near a ${\bf{q_{0}}}$=($\frac{1}{2}$,0) to gapless fluctuations at ${\bf{q_{0}}}$=($\frac{1}{2}$,$\frac{1}{2}$).  We also observe a filling in of low energy spectral weight with increased interstitial iron doping.  These results point to the charge doping from interstitial iron playing a key role in the electronic and magnetic properties of Fe$_{1+x}$Te$_{1-y}$Se$_{y}$.  

\section{Experiment}

A powder sample of nominal composition Fe$_{1.05}$Te$_{0.7}$Se$_{0.3}$ was synthesized by a solid state reaction of the constituent elements at 700 $^{\circ}$C under vacuum.   Samples of varying interstitial iron were synthesized by exposing the powders to various levels of I$_{2}$, as outlined in Ref. \onlinecite{Rodriguez10:132}.      The interstitial iron concentration was then determined through the use of both x-ray and neutron diffraction as outlined in Ref. \onlinecite{Rodriguez11:2}, which discusses a crystallographic study on the same materials.  The masses, interstitial iron concentration, and lattice parameters are are listed in Table \ref{table:sample}.  Attempts to apply this technique to single crystalline materials has currently not been successful or resulted in large inhomogeneities.

Neutron inelastic scattering results were performed using the Disk Chopper Spectrometer (DCS) located at the NIST Center for Neutron Research.  The powder samples were closed in an Helium flow cryostat and confined within cylindrical sample cans (of radius $R$) such $\mu R$ $\textless$ 0.1 (where $\mu$ is the absorption factor and the correction factors for a cylinder are tabulated in Ref. \onlinecite{Dwiggins75:31,tables}).  We have therefore not included any correction for multiple scattering or absorption of the neutron beam.  An incident energy of 14 meV was chosen such that an elastic ($\hbar\omega$=0) energy resolution (full-width) of 0.92 meV was obtained.  The DCS instrument consists of 913 detectors with active dimensions in and normal to the scattering plane of $\approx$ 3.1 and 40 cm, respectively, covering scattering angles from $2\theta$=5$^{\circ}$ to 140$^{\circ}$.  Further details of the instrument can be found elsewhere.~\cite{Copley03:292}

\begin{table}[ht]
\caption{A summary of the sample characteristics used in this study.}
\begin{tabular}{c c c c }
\hline\hline
 $Sample$ & $a$(\AA) & $c$ (\AA) & $mass (g)$\\
 \hline\hline
 Fe$_{1.009(3)}$Se$_{0.3}$Te$_{0.7}$ & 3.8047(1)  & 6.0676(2) & 5.1(1)  \\
 Fe$_{1.018(2)}$Se$_{0.3}$Te$_{0.7}$ & 3.80220(3) & 6.0750(1) & 5.4(1) \\
 Fe$_{1.033(2)}$Se$_{0.3}$Te$_{0.7}$ & 3.80112(3) & 6.0787(1) & 2.9(1) \\
 Fe$_{1.048(2)}$Se$_{0.3}$Te$_{0.7}$ & 3.80224(2) & 6.0756(1) & 8.7(1) \\
 \hline
\end{tabular}
\label{table:sample}
\end{table}

Interpretation of powder data is complicated by the fact that the spin fluctuations are averaged over momentum transfer.  It is therefore important to subtract off contributions resulting from any phonons and also any temperature independent background.  In the next section, we discuss how we have estimated these contributions by imposing detailed balance.

\section{Data analysis and background subtraction}

The temperature independent background resulting from elastic scattering leaking through into the inelastic channels owing to the finite resolution function and also instrumental effects can be determined through the fact that inelastic scattering must obey detailed balance.    This technique has been outlined in detail and applied to a polymer quantum magnet in Ref. \onlinecite{Hong06:74}.  To a good approximation, for a fixed wave vector and energy transfer, the neutron energy gain (negative energy transfer) and energy loss (positive energy transfer) are related by the following expression for a fixed wave vector transfer.

\begin{eqnarray}
\label{detailed_balance} 
I_{mes}(+|E|,T)=B_{1}(E)+S(|E|,T) \\ \nonumber
I_{mes}(-|E|,T)=B_{2}(-|E|)+S(|E|,T) e^{-{{E} \over {kT}}} 
\end{eqnarray}

\noindent In this expression, $B_{1}$ and $B_{2}$ are temperature independent background points and $S(|E|,T)$ is the signal which is a summation of the phonon and magnon scattering.  The factor $e^{-{{E} \over {kT}}}$ is the Boltzmann factor.  An assumption in this analysis is that the resolution function does not change substantially over the energy range investigated.  It can be seen with at least two different temperatures, the temperature independent background $B_{1}$ and $B_{2}$ can be determined.

\begin{figure}
\includegraphics[width=8.5cm]{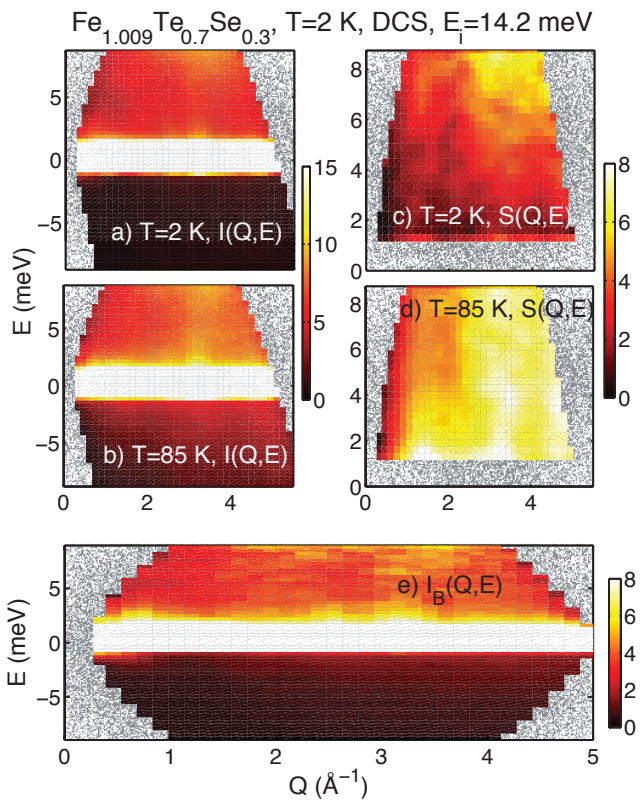}
\caption{\label{extract_sqw} (Color online) $a)$ and $b)$ display the measured intensity on DCS at T=2 and 85 K.  $c)$ and $d)$ illustrate S($Q$,E) at these temperatures and the temperature independent background is presented in $e)$.}
\end{figure}

An example of this background subtraction is presented in Fig. \ref{extract_sqw} which illustrates the measured intensity spectrum for Fe$_{1.009}$Te$_{0.7}$Sr$_{0.3}$ at 2 and 85 K in panels $a)$ and $b)$.   The extracted S(Q,E), based upon detailed balance, is plotted in panels $c)$ and $d)$.  The temperature independent background is shown in $e)$.  This analysis was performed independently for all iron concentrations studied.

\begin{figure}
\includegraphics[width=8.5cm]{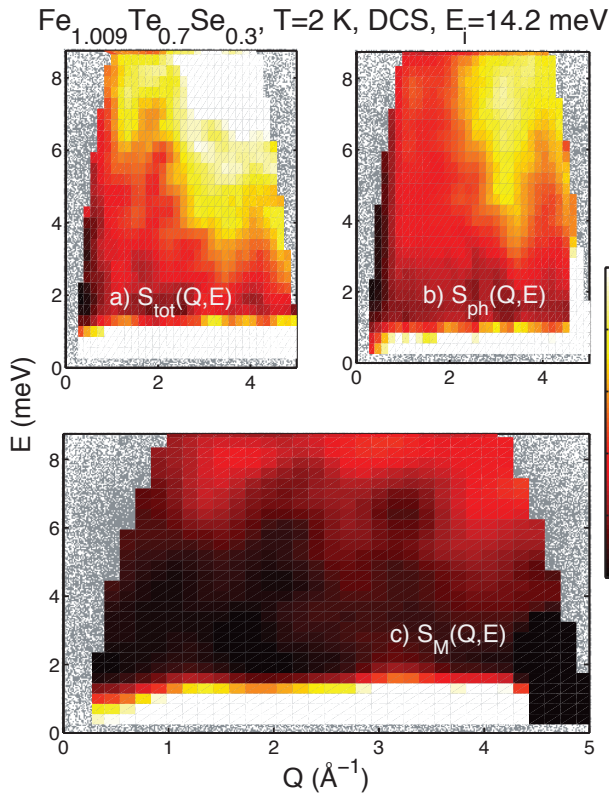}
\caption{\label{subtract} (Color online) $a)$ is the total $S(Q,E)$ derived using detailed balance and after background subtraction for Fe$_{1.009}$,Te$_{0.7}$Se$_{0.3}$.  $b)$ is the phonon background obtained from the method described in the text.  $c)$ is the subtraction illustrating the magnetic scattering. The analysis fails at the largest values of $Q$ and lowest energies.}
\end{figure}

After subtracting the temperature independent background we have used the 150 K and 200 K data for the Fe$_{1.009}$Te$_{0.7}$Sr$_{0.3}$ as an estimate for the phonon background.  We chose this sample as it displays the least amount of spectral weight at low-energies and at low momentum transfers indicating that it displays the smallest amount of magnetic scattering amongst the four samples studied.   This is corroborated by previously reported single crystal work which has shown that the magnetic scattering is smeared out in energy and momentum at high temperatures.  The phonon background was then determined as follows.

\begin{eqnarray}
\label{chi_phonon} 
\chi''_{phonon}(Q,E)=...\\ \nonumber
{1\over2}\left({S_{150 K}(Q,E) \over {e^{E/kT=150K}+1}}+{S_{200 K}(Q,E) \over {e^{E/kT=200K}+1}} \right)
\end{eqnarray}

\noindent The respective denominators are the bose factors for the measurements using the relation $S(Q,E) \propto [n(E)+1] \chi''(Q,E)$.  Having derived $\chi''_{phonon}(Q,E)$ we can then derive $S_{ph}(Q,E)$ for the required temperature being measured.  The magnetic component of $S(Q,E)$, denoted $S_{M}(Q,E)$, was derived from $S_{M}(Q,E)=S(Q,E)-S_{ph}(Q,E)$.

The results of this subtraction for the Fe$_{1.009(3)}$Se$_{0.3}$Te$_{0.7}$ sample are illustrated in Fig. \ref{subtract} at 2 K, within the superconducting state.  The subtraction seems to work well at low momentum transfers below 3 \AA$^{-1}$, the strong phonon scattering at large momentum transfers ($\approx$ 4 \AA$^{-1}$) may be associated with anharmonic effects.  We will concentrate our investigation on the dynamics below $Q=2.5$ \AA$^{-1}$ for the remainder of this paper.

To put all of the data on a common scale from which absolute units could be obtained, we have normalized all the intensity in each sample to the integrated intensity over the range $Q$=[3,4] \AA$^{-1}$ and E= [5,8] meV where the spectral weight is assumed to be dominated by phonon scattering and should depend weakly on small amounts of interstitial iron doping.   A comparison of absolute integrated intensity is discussed later in the paper.

\section{Results}

\begin{figure}
\includegraphics[width=8.5cm]{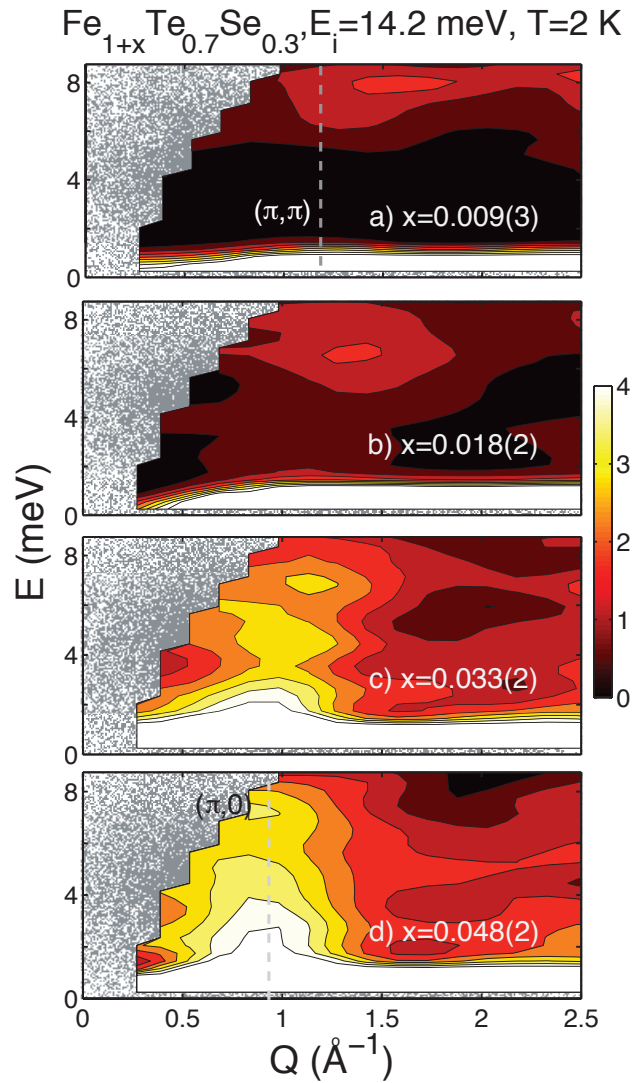}
\caption{\label{summary} (Color online) $a-d)$ illustrates the contours of magnetic intensity derived for the samples described in the text. The momentum positions of ${\bf{q}}=(\pi,0)$ and ${\bf{q}}=(\pi,\pi)$ are indicated by dashed lines.}
\end{figure}

A summary of the extracted magnetic intensity normalized to a common intensity scale is illustrated in Fig. \ref{summary} at 2 K for all the interstitial iron concentrations studied.   For low interstitial iron concentrations ($x$=0.009(3)) with the maximum superconducting volume fraction, the magnetic excitations are gapped with a value of $\approx$ 7 meV.  This matches the resonance energy observed previously in single crystal samples with T$_{c} \approx$ 14 K.  On doping with interstitial iron the magnetic fluctuations fill in at lower energy and move to lower momentum transfers.  At $x$=0.048(2), where the superconducting volume fraction has been suppressed to zero (Fig. \ref{phase_diagram}), the magnetic fluctuations are gapless and are centered in momentum transfer at the $(\pi,\pi)$ position.  

In this section, we investigate the energy and momentum evolution of the magnetic fluctuations as a function of interstitial iron.  We will demonstrate that while doping interstitial iron suppresses superconductivity, it also lowers the energy scale of the spin fluctuations and shifts the spectral weight in momentum from the $(\pi,\pi)$ position in superconducting samples to the $(\pi,0)$ point.  We first investigate the temperature dependence of the spin fluctuations in the $x$=0.009(3) sample and compare it with single crystal studies.

\begin{figure}
\includegraphics[width=8.5cm]{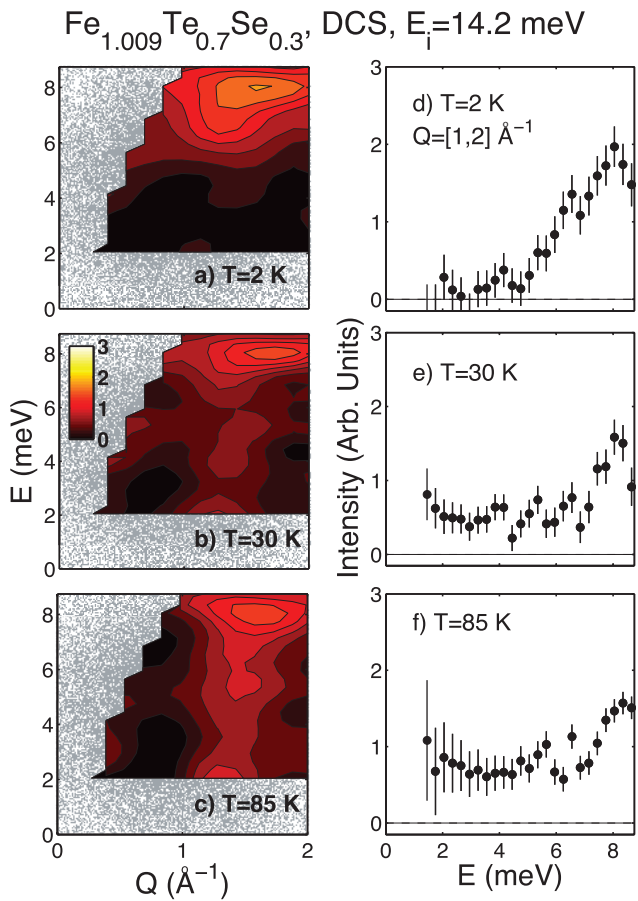}
\caption{\label{temp_evolve} (Color online) False-color contour maps of the temperature evolution of the spin fluctuations are illustrated in the left hand panels.  Momentum integrating energy scans are presented for the same temperatures in the right-hand panels.}
\end{figure}

\subsection{Temperature dependence of the spin fluctuations in Fe$_{1.009(3)}$Te$_{0.7}$Se$_{0.3}$}

The background and phonon subtracted intensities maps for Fe$_{1.009(3)}$Te$_{0.7}$Se$_{0.3}$ at several temperatures is illustrated in Fig. \ref{temp_evolve}.  This interstitial iron concentration contains a large superconducting volume fraction with an onset of superconductivity at T$_{c}$=14 K.  As described above, the low temperature powder averaged spectrum is qualitatively consistent with the single crystal experiments at $y \approx 0.5$ which display optimal superconductivity.  The powder data at T=2 K presented in Fig. \ref{temp_evolve} $a)$ and $d)$ displays a clear gap of $\approx$ 7 meV, as expected for superconducting samples.  

On increasing temperature the low-energy spectrum fills in with a large amount of spectral weight still present above 7 meV, though reduced from T=2 K result.  While this result is surprising given expectations on the 7 meV being the resonance directly correlated with superconductivity, this result on powder samples is consistent with recent single crystal experiments which have observed little change in the spectral weight present above the low temperature spin-gap with temperature.~\cite{Xu11:84}  These results have been interpreted as strong evidence for the role of local magnetism in the iron telluride superconductors.  The low-energy spectral weight present below the low temperature spin gap increases with temperature and at 85 K (Fig. \ref{phase_diagram} (c) and (f)) the magnetic fluctuations are nearly evenly distributed in spectral weight over the entire energy range probed. 

These results are consistent with the temperature dependence probed in single crystal samples and confirm the presence of spectral weight above the spin-gap energy.  It also confirms the validity of the phonon and background subtraction performed and discussed in the previous section and demonstrates that the intensity obtained from the subtraction is purely magnetic.    We now investigate the momentum and energy dependence of the magnetic spectral weight as a function of interstitial iron doping.

\subsection{First moment and the Hohenberg-Brinkman sum rule} 

\begin{figure}
\includegraphics[width=7.5cm]{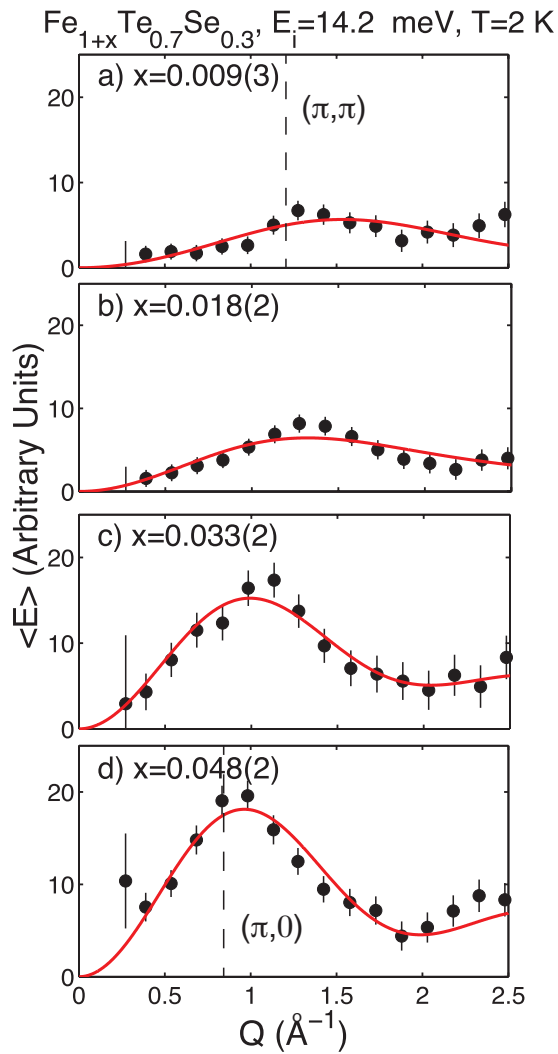}
\caption{\label{first_moment_figure} (Color online) The first momentum in energy as a function of momentum transfer is illustrated for the interstitial iron concentrations investigated.  The solid curves are fits to the Hohenberg-Brinkman sum rule described in the text.}
\end{figure}

To study the spatial correlations as a function of interstitial iron we have calculated the first-moment in energy by integrating the data over all energies studied.  The first moment is formally defined as,

\begin{eqnarray}
\label{first_moment} 
\langle E ({\bf{Q}}) \rangle = \int_{-\infty}^{\infty} dE \ E S({\bf{Q}},E)
\end{eqnarray}

\noindent where $S({\bf{Q}},E)$ is the measured magnetic structure factor (measured after background and phonon subtraction), and $E$ and ${\bf{Q}}$ are the energy and momentum transfers, respectively.  

The experimental result is illustrated in Fig. \ref{first_moment_figure} for all interstitial iron concentrations studied at T=2 K.   The integral in energy was performed over the range E=[2, 8.75] meV in energy transfer.  While the first moment should be defined in terms of the integral over all of the spectral weight over all energy transfers, this is not practical in the case of the iron tellurides where the spin excitations have been shown to extend up to more than $\approx$200 meV.~\cite{Lumsden10:6,Lipscombe11:106}.  Also, we expect the effect of interstitial iron and superconductivity will affect the low-energy excitations below $\approx$ 10 meV as confirmed in single crystal studies on superconducting concentrations.~\cite{Argyriou10:81}  This has also been observed through temperature dependent studies of the spin fluctuations which have reported little change in the spectral weight above the resonance energy ($\approx$ 7 meV), yet large changes at lower energy transfers. ~\cite{Xu11:84} 

From the first moment, we can derive several model independent results from data.  The first moment is described by sum rules given by Hohenberg and Brinkman.~\cite{Hohenberg94:10}   For the case of isotropic exchange, the first moment sum was found to have the following form for a single crystal.

\begin{eqnarray}
\label{first_moment} 
\langle E ({\bf{Q}}) \rangle =-{1\over 3} {1 \over N} {\sum_{{\bf{r}},{\bf{d}}}} J_{{\bf{d}}} \langle  {\bf{S}}_{\bf{r}} \cdot {\bf{S}}_{{\bf{r}}+{\bf{d}}}  \rangle (1-\cos({\bf{Q}}\cdot{\bf{d}}))
\end{eqnarray}

\noindent The powder average of the first moment can be written as follows,~\cite{Hammar98:57}

\begin{eqnarray}
\label{first_moment_powder} 
\langle E ({\bf{Q}}) \rangle \propto |f(Q)|^{2}  {\sum_{{\bf{r}},{\bf{d}}}} J_{{\bf{d}}} \langle  {\bf{S}}_{\bf{r}} \cdot {\bf{S}}_{{\bf{r}}+{\bf{d}}} \rangle \left( 1- {{\sin(Qd)} \over {Qd}} \right).
\end{eqnarray}

\noindent where $f(Q)$ is the magnetic form factor, $J_{\bf{d}}$ the exchange constant, and  $\bf{S_r}$ the spin moment on the iron at position $\bf{r}$, and $d$ the interatomic distance.  This expression allows a model independent means of parameterizing the data.  The use of the first moment sum rules also allows microscopic information to be obtained on which spins are strongly correlated.

The solid curves in Fig. \ref{first_moment} are the result of fits to Eqn. \ref{first_moment_powder} taking the first two terms in the sum for nearest neighbors ($d_{1}=2.69 \AA$) and next-nearest neighbor interactions ($d_{2}=3.81 \AA$).   With the distances ($d_{i=1,2}$) fixed from the low temperature crystallographic data, the measured first moment can be fit to a model with two parameters ($J_{{\bf{d_{i}}}} \langle  {\bf{S}}_{\bf{r}} \cdot {\bf{S}}_{{\bf{r}}+{\bf{d_{i}}}} \rangle$) which represent the strength of the nearest and next nearest neighbor correlations.  

The $x$=0.009(3) sample displays a weakly peaked spectrum as a function of momentum transfer.  While the data is statistically limited, it is consistent  with strong correlations resulting from nearest neighbors.  As the interstitial iron concentrations is increased, this peak moves to lower momentum transfers where the $x$=0.048(2) sample requires strong correlations from both both nearest and next nearest neighbors to accurately describe the data.  These results illustrate that as the interstitial iron is increased, the magnetic fluctuations move from being peaked at a momentum transfer of $(\pi,\pi)$ to the $(\pi,0)$ point.  While this result has been reported previously for selenium doping, here we demonstrate that the same result can be reproduced with charge doping with interstitial iron.

\subsection{Energy dependence as a function of interstitial iron}  

\begin{figure}
\includegraphics[width=7.5cm]{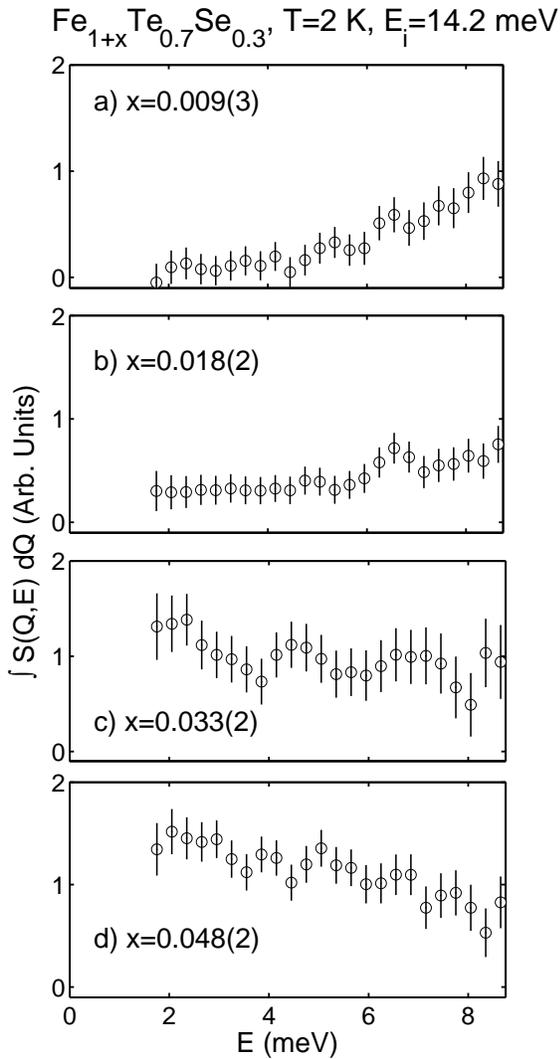}
\caption{\label{Q_integrate} The momentum integrated intensity is plotted at T= 2 K.}
\end{figure}

The momentum integrated intensity over the range of ${\bf{Q}}=[0,2.35] \AA ^{-1}$ for all interstitial iron concentrations studied is presented in Fig. \ref{Q_integrate}.   The data illustrates that with increasing interstitial iron more spectral weight gathers at low-energy transfers below the resonance energy of $\approx$ 7 meV.   Fig \ref{Q_integrate} (a) shows that for the sample with the largest superconducting volume fraction, the magnetic spectral weight exists at energy transfers larger than $\approx$ 7 meV.  On increasing the amount of interstitial iron (illustrated in Fig. \ref{Q_integrate} (b)-(d)), the spectrum gradually fills in at energy transfers below the spin-gap found in the most superconducting sample.  For the largest interstitial concentration studied of $x$=0.048(2), no sign of a spin-gap is observable in the data and the spectrum smoothly varies with energy.  

These results indicate that the amount of spectral weight at low-energies is directly tied to the superconducting volume fraction in the iron telluride superconductors.   With interstitial iron doping we do not observe a continuous drop in the superconducting transition temperature but rather a decrease in the volume fraction (as illustrated in Fig. \ref{phase_diagram} (b)).  While the volume fraction decreases (with increasing interstitial iron concentration), the characteristic energy scale is tuned to low energies.  These results indicate that the magnetic energy scale can be tuned through charge doping with interstitial iron, a similar conclusion derived from a single crystalline study on the parent Fe$_{1+x}$Te which investigated the low-energy fluctuations as a function of interstitial iron concentration.

\subsection{Absolute units and spectral weight}  

Having discussed how the magnetic inelastic spectrum varies in momentum and energy as a function of interstitial iron concentration, we now estimate the absolute total moment which resides in our experimental window.  To put the data on an absolute scale, we have calibrated the spectrometer by performing a scan through the incoherent elastic line over the range $Q=[1.7,1.8]$ \AA$^{-1}$ ,where no strong coherent Bragg scattering is present.  The cross section can then be written as,

\begin{eqnarray}
\label{incoh_line} 
{d^{2}\sigma \over d\Omega dE}= A \sum_{i} b_{i}^{2} \delta(E),
\end{eqnarray}

\noindent where $A$ is a calibration constant which depends on the spectrometer count rate and also the number of atoms in the unit cell, $\delta(E)$ is the Dirac delta function and $b_{i}$ are the incoherent scattering lengths of the constituent atoms.  Assuming that the scattering is dominated by the incoherent scattering lengths for Fe, Te and Se, we can then calculate $A$ by integrating the elastic line.

To compare our results with those of other groups and with the total moment expected by sum rules, we take the magnetic cross section to be defined as follows,

\begin{eqnarray}
\label{incoh_line} 
{d^{2}\sigma \over d\Omega dE}= A {{(\gamma r_{0})^{2}} \over 4}  g^{2} f(Q)^{2} 2 \times S(Q,E),
\end{eqnarray}

\noindent where $\left({{\gamma r_{0}^{2}} \over 4} \right)$ is 73 mbarns, $g$ is the Lande factor, $f(Q)$ is the magnetic form factor for Fe$^{2+}$, and $S(Q,E)$ is the magnetic powder averaged scattering function.  $S(Q,E)$ is governed by the following sum rule for a localized system,

\begin{eqnarray}
\label{integral} 
I=\int d^{3}Q \int dE S(Q,E)/ \int d^{3}Q = S(S+1).
\end{eqnarray}

\noindent Assuming a localized moment scenario with $g=2$, we obtain an integrated value over E=[2.0, 8.75] meV and Q=[0, 2.35] \AA$^{-1}$ of I=0.007(2) and 0.012 (2) for x=0.009(3) and x=0.048(2), respectively.  These are a small fraction of the expected value of 6 (taking $S$=2)  predicted from the total moment sum rule owing to the large bandwidth of the excitations.  We note that Ref. \onlinecite{Turner09:80} argues for S=1 which would suggest a total moment of 2.  In either case, our calibration does indicate that the amount of spectral weight residing in the resonant excitation may amount to no more than a few percent of the total spectral weight expected based on a localized moment picture.  Whether this is consistent with expectations was a matter of a debate in the cuprate superconductors.~\cite{Kee02:88,Abanov02:89}  The presence of such little spectral weight in the resonance energy maybe consistent with the fact that resonant peak is localized both in momentum and energy.

These values are broadly consistent with absolute values obtained in the cuprate superconductors over a similar range and therefore suggests the values stated here do have some physical credence.~\cite{Stock08:77,Stock04:69}  These values are comparable to those obtained by other groups which have probed the dynamics in the superconducting phase.~\cite{Xu11:84}  We note that the calibration method here does assume that the elastic cross section is dominated by the incoherent cross section for Fe$_{1+x}$Te$_{0.7}$Se$_{0.3}$.  It will be important to check this calibration to internal phonons as well as an external vanadium in single crystal samples.  The analysis also assumes that orbital fluctuations are weak which might not be the case.~\cite{Lee10:81}

\section{Conclusions}

\begin{figure}
\includegraphics[width=7.5cm]{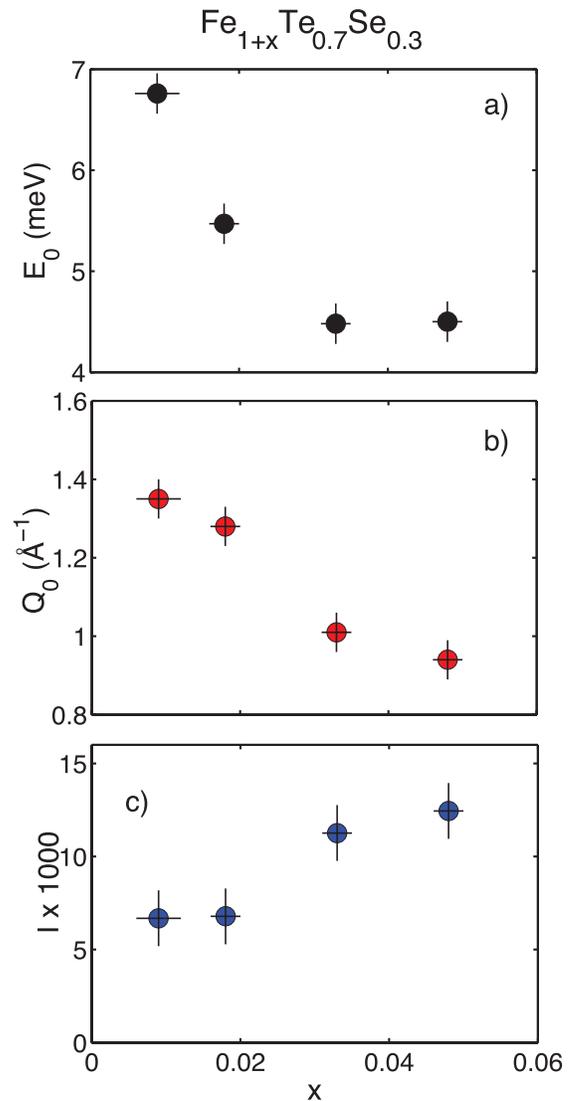}
\caption{\label{param} (Color online) (a) Peak position of the magnetic spectrum in momentum. (b) Mean energy position as defined in the text.  (c) Total integrated intensity in energy and momentum as a function of interstitial iron concentration.  All of the data is presented for T=2 K.}
\end{figure}

A summary of the characteristic energy, momentum, and integrated intensity as a function of interstitial iron concentration is provided in Fig. \ref{param}.  The mean energy $E_{0}$ was defined as $\int dE \ E S(E) / \int dE S(E)$.  The characteristic momentum value $Q_{0}$ has obtained by fitting a gaussian to the lowest angle peak displayed in  Fig. \ref{first_moment_figure}.  The integrated magnetic spectral weight defined in Eq. \ref{integral} is plotted in Fig. \ref{param} (c).   The peak in momentum shifts from the ${\bf{Q}}=(\pi,\pi)$ point to the ${\bf{Q}}=(\pi,0)$ as illustrated in Fig. \ref{param} (a).  Simultaneously, the average energy position decreases (Fig. \ref{param} (b)) until it is equal to the average energy transfer probed by the experiment indicating the presence of gapless two dimensional fluctuations.  These results imply that spin fluctuations near ${\bf{Q}}=(\pi,0)$ destroy superconductivity.  Therefore, having spin fluctuations which match the nesting wave vector of  ${\bf{Q}}=(\pi,\pi)$ seems to be beneficial for superconductivity.

A recent study on Fe$_{1.01}$Te$_{0.7}$Se$_{0.3}$ reported the coexistence of fluctuations near ${\bf{Q}}=(\pi,\pi)$ and ${\bf{Q}}=(\pi,0)$ in the same sample.~\cite{Chi11:xx}  These results indicate that the shift of fluctuations in wavevector is discontinuous and that dependence of interstitial iron dependence of $Q_{0}$ displayed in Fig. \ref{param} $b)$ is a discontinuous one from ${\bf{Q}}=(\pi,\pi)$ (Q=1.2$\AA^{-1}$) to ${\bf{Q}}=(\pi,0)$ (Q=0.8$\AA^{-1}$).  Our results indicate that the coexistence of the fluctuations maybe removed through the extraction of interstitial iron and that change in wavevector occurs over a very narrow region of interstitial iron.  Our data seems to suggest that the change in wavevector may indeed occur at $x\sim0.03$.  

The dependence of the integrated spectral weight presented in Fig. \ref{param} (c) is particularly interesting as it varies only by a factor of $\approx$ 2 while changing the amount of interstitial iron dramatically.  The amount of spectral weight at low energies does not depend directly on the interstitial iron concentration, but the change in spectral weight can be accounted for by the fact that the excitations are gapless for large interstitial iron concentrations and gapped for smaller values.  Our analysis implies that the spectral weight is not induced with interstitial iron concentration, but rather shifted both in momentum and energy.    Therefore, our results seem to imply a localized moment picture with possibly the exchange interactions changing with interstitial iron and hence the characteristic wave vector of the spin fluctuations.

Two key results have been reported previously for the iron telluride superconductors as a function of selenium doping on the anion site.  First, a spin gap or resonance develops with an energy of $\approx$ 7 meV.  Second, the magnetic scattering shifts from the ${\bf{Q}}=(\pi,0)$ point to the ${\bf{Q}}=(\pi,\pi)$ position in momentum transfer.   We observe in this experiment that these two key results for the spin fluctuations in the iron telluride superconductors can be reproduced by tuning the interstitial iron concentration over a narrow range.  These results imply the substitution on the anion site is not the only driving point for superconductivity; the doping of interstitial iron, which would affect the oxidation state of the in-plane iron, is also important.  Indeed, the role of selenium doping appears to be to facilitate smaller concentrations of interstitial iron, which was demonstrated clearly with a study involving doping sulphur instead of selenium.~\cite{Zajdel10:132}  This study illustrates the need to characterize the amount of interstitial iron in a particular sample and likely suggests that the phase diagram presented in Fig. \ref{phase_diagram} is also controlled by the amount of interstitial iron in addition to selenium.  

It is important to note that none of the samples studied here displayed elastic (E=0 meV) magnetic correlations indicative of long-range magnetic order.  It is therefore not clear whether the role of selenium is also to suppress magnetic order while also facilitating lower interstitial iron concentrations.   It will be interesting to try and reduce the interstitial iron of Fe$_{1+x}$Te to study the magnetic order and existence of a superconducting volume fraction.  Current investigations have only be able to reduce the interstitial iron to as a low as $x \approx$0.04 in the parent material.\cite{Rodriguez10:132}

In summary, we have demonstrated the effects of interstitial iron doping on the spin fluctuations and superconductivity in Fe$_{1+x}$Te$_{0.7}$Se$_{0.3}$.  We have shown that interstitial iron shifts the characteristic wavevector from ${\bf{Q}}=(\pi,\pi)$ to ${\bf{Q}}=(\pi,0)$ with increasing interstitial iron concentration.  Simultaneously, the superconducting volume fraction decreases.  We also observe that the total integrated moment at low energies does not vary linearly with interstitial iron concentration consistent with a localized moment scenario with a redistribution of spectra weight. 


\end{document}